%% file: main.tex
\documentclass[10pt,conference]{IEEEtran}
\IEEEoverridecommandlockouts

\usepackage{subcaption}

\usepackage{scalerel}
\usepackage{multirow} 
\usepackage{booktabs}
\usepackage[table]{xcolor}
\usepackage[sort&compress, numbers]{natbib}
\usepackage{amsmath,amssymb,amsfonts}
\usepackage{algorithmic}
\usepackage{graphicx}

\usepackage{float}\usepackage{textcomp}
\usepackage{xcolor}
\usepackage{flushend}
\def\BibTeX{{\rm B\kern-.05em{\sc i\kern-.025em b}\kern-.08em
    T\kern-.1667em\lower.7ex\hbox{E}\kern-.125emX}}

\usepackage{tcolorbox}
\usepackage[inline]{enumitem}
\usepackage{tabularx} 

\usepackage{xcolor}  
\definecolor{codegreen}{rgb}{0,0.6,0}
\definecolor{codegray}{rgb}{0.5,0.5,0.5}
\definecolor{codepurple}{rgb}{0.58,0,0.82}
\definecolor{backcolour}{rgb}{0.95,0.95,0.92}

\usepackage{listings}
\lstdefinestyle{mystyle}{
  aboveskip=3mm,
  showstringspaces=false,
  columns=flexible,
  numbers=none,
  backgroundcolor=\color{backcolour},
  commentstyle=\color{codegreen},
 keywordstyle=\color{magenta},
    numberstyle=\tiny\color{codegray},
    stringstyle=\color{codepurple},
    basicstyle=\scriptsize\ttfamily,
    breakatwhitespace=false,         
    breaklines=true,                 
    captionpos=b,                    
    keepspaces=false,                 
    numbersep=5pt,                  
    showspaces=false,                
    showstringspaces=false,
    showtabs=false,                  
    tabsize=2,
    escapeinside=``
}
\usepackage{natbib}

\usepackage{hyperref}

\newcommand{\cdiff}{$C_{\textsc{diff}}$}
\newcommand{\rnl}{$R_{\textsc{nl}}$}

\newcommand{\tvalid}{\texttt{valid}}
\newcommand{\tnoisy}{\texttt{noisy}}

\newcommand{\cleangpt}{$\textsc{Cleaned}_{\textsc{GPT-3.5}}$}
\newcommand{\cleanllama}{$\textsc{Cleaned}_{\textsc{Llama3}}$}

\newcommand{\controlgpt}{$\textsc{Controlled}_{\textsc{GPT-3.5}}$}
\newcommand{\controlllama}{$\textsc{Controlled}_{\textsc{Llama3}}$}

\newcommand{\pdef}{$\text{P}_{\textsc{Definition}}$}
\newcommand{\paux}{$\text{P}_{\textsc{Auxiliary}}$}

\begin{document}


\title{Too Noisy To Learn: Enhancing Data Quality for Code Review Comment Generation}

 \author{\IEEEauthorblockN{Chunhua Liu}
 \IEEEauthorblockA{
 \textit{The University of Melbourne}\\
 chunhua.liu1@unimelb.edu.au}
  \and
 \IEEEauthorblockN{Hong Yi Lin}
 \IEEEauthorblockA{
 \textit{The University of Melbourne}\\
 holin2@student.unimelb.edu.au}
 \and
 \IEEEauthorblockN{Patanamon Thongtanunam}
 \IEEEauthorblockA{
 \textit{The University of Melbourne}\\
 patanamon.t@unimelb.edu.au}
 }

\maketitle

\begin{abstract}
Code review is an important practice in software development, yet it is time-consuming and requires substantial effort. While open-source datasets have been used to train neural models for automating code review tasks, including review comment generation, these datasets contain a significant amount of noisy comments (e.g., vague or non-actionable feedback) that persist despite cleaning methods using heuristics and machine learning approaches. Such remaining noise may lead models to generate low-quality review comments, yet removing them requires a complex semantic understanding of both code changes and natural language comments. In this paper, we investigate the impact of such noise on review comment generation and propose a novel approach using large language models (LLMs) to further clean these datasets. Based on an empirical study on a large-scale code review dataset, our LLM-based approach achieves 66-85\% precision in detecting valid comments. Using the predicted valid comments to fine-tune the state-of-the-art code review models (cleaned models) can generate review comments that are 13.0\% - 12.4\% more similar to valid human-written comments than the original models. We also find that the cleaned models can generate more informative and relevant comments than the original models. Our findings underscore the critical impact of dataset quality on the performance of review comment generation. We advocate for further research into cleaning training data to enhance the practical utility and quality of automated code review.

\end{abstract}


\section{Introduction}

Code review is a critical practice in software development, providing multiple benefits such as identifying logic errors~\cite{Bacchelli-etal-2013-Expectations,Rigby-etal-2013-Convergent}. At the core of this process is the reviewers' comments on code changes, which serve as the primary means for providing feedback and suggestions. 
Despite its advantages, the process is often time-consuming and requires substantial effort from reviewers~\cite{Bosu-etal-2017-Process}. To alleviate this burden, recent research has focused on automatically generating review comments by training neural models on large-scale datasets mined from open-source platforms such as GitHub and Gerrit. 
While these models show promise in generating code review comments by training neural models on large-scale datasets~\cite{Tufano2022PretrainedModels,Li2022CodeReviewer,CCT5}, the gap between model-generated and human-written comments remains significant, limiting their practical use. 

\input{tables/example_reviewer_comments_valid_noisy}

Despite the availability of large-scale datasets, the quality of review comments varies due to various factors such as reviewer experience, limited review time~\cite{Oleksii-etal-2016-CodeReviewQuality, Lin-etal-2024-Improving}, diverse communicative intentions~\cite{ebert2018communicative}, and different standards for code review across different projects. Prior work has attempted to clean these datasets by removing bots and filtering out noisy comments using heuristics~\cite{tufano2021towards, Tufano2022PretrainedModels}. More recently, Li et. al. have attempted to clean their widely-used CodeReviewer benchmark using heuristic rules and a machine learning classifier (SVM)~\cite{Li2022CodeReviewer}. 
Yet concerns about data quality persist. A recent work by Tufano et al.~\cite{tufano2024code} revealed that a substantial proportion (32\%) of comments in the test set of this widely-used dataset still contains noise. The noise includes vague, difficult-to-understand comments, and comments that merely seek clarification rather than suggest an improvement to the code (see Fig. \ref{fig:code_reviews_stacked} Top).


We argue that such noisy comments may also exist in the training set. Neural models trained on noisy datasets inevitably internalize and potentially propagate low-quality reviews into the generated review comments. Such poor reviews can be considered as less useful in practical settings~\cite{bosu2015characteristics,Oleksii-etal-2016-CodeReviewQuality, turzo2024makes} or even negatively impact software quality~\cite{Mcintosh-etal-2016-ImpactCodeReiew}. To develop models that generate high-quality review comments—providing clear suggestions for code improvements that assist code authors and automated code refinement—it is crucial to improve the quality of the training data.



Identifying and removing noisy comments remaining in the code review datasets presents significant challenges due to the complexity of the task. It requires an understanding of both the technical context of code changes and the review comments written in natural language. Therefore, interpreting code review comments can be highly ambiguous and inference-heavy~\cite{ebert2018communicative,Ebert-etal-2021-ConfusionCodeReview}. Simple heuristics such as keywords matching and sentence length used in prior works~\cite{tufano2021towards,Tufano2022PretrainedModels} fall short in analyzing these complexities and nuances in the review comments, urging the need for a more semantic and context-aware method to effectively clean code review datasets.

To address these challenges, this work explores a novel approach to improve the quality of the review datasets and studies the impact of noise on the performance of automated review comment generation.  
As large language models (LLMs) have shown promising performance in understanding both code and natural language \cite{chen2021evaluating,hendrycks2020measuring}, and potential data annotation tasks in other domains \cite{Gilardi-etal-2023-ChatGPTAnnotation,ding-etal-2023-gptAnnotator}, we investigate the feasibility of using LLMs to classify noisy and valid comments, then retain the valid ones. Finally, we evaluate the performance of the state-of-the-art models (i.e., CodeReviewer~\cite{Li2022CodeReviewer} and CodeT5~\cite{wang-etal-2021-codet5}) trained on the cleaned datasets.


Through an empirical study on the widely-used code review benchmark (CodeReviewer~\cite{Li2022CodeReviewer}), we find that
only 64\% of the sampled comments in the training set of CodeReviewer benchmark are valid.  Our approach using LLMs achieved a precision of 66\% - 85\% in identifying valid comments and achieved a recall of 51\% - 89\% in identifying noisy comments.
These results suggest that 
the proportion of valid comments in the dataset can be improved from 64\% to 85\% by LLMs. By retaining only the valid comments predicted by LLMs (i.e., cleaned datasets), the training size is 25\% - 66\% smaller than the original dataset.
Nonetheless, the smaller data did not negatively impact the performance of comment generation models.   
Instead, the models fine-tuned on the cleaned datasets achieved BLEU-4 scores 7.5\% - 13\% higher than those trained on the original dataset, with a 12.4\% - 13.0\% increase specifically on valid comments in test sets.
Moreover, we also found that the quality of comments generated from the cleaned models is significantly improved, with up to a 24\% increase in information score and an 11\% increase in relevance score. These results highlight that valid and noisy review comments can be detected by LLMs and cleaning review data can improve the performance of the review comment generation models.

\textbf{Novelty and Contribution.} To the best of our knowledge, we are the first to
\begin{enumerate*} 
    \item present an automated approach to clean the large-scale review dataset using LLMs,
    \item demonstrate the capability of LLMs to automatically classify valid and noisy review comments, and
    \item highlight the impact of data quality on the performance of automated review comment generation, \item demonstrate an improvement in model performance and comment quality with a cleaned dataset despite its significantly smaller size than the original data (e.g., 117K vs 39K), \item employ a semi-automated method to approximate the quality of generated review comments at scale.
\end{enumerate*}

\textbf{Open Science.}
To facilitate reproducibility and future work, we provide a replication package that includes the cleaned datasets, experimental results, and scripts.\footnote{\url{https://zenodo.org/records/13150598}}

\textbf{Paper Organisation.}
The remainder of this paper is structured as follows: 
Section~\ref{sec:related_work} discussed related work. 
Section~\ref{sec:study_design} outlines an overview of the study design. 
Sections~\ref{sec:identifying_noisy_comments_rq1} - \ref{sec:noisy_impact_on_generated_comment_quality_rq3} present our study approaches and
results of each research question. 
Section~\ref{sec:discussion} discusses the findings. Section~\ref{sec:threats_to_validity} addresses potential threats to the validity. Finally, Section~\ref{sec:conclusion} draws a conclusion.

\section{Background \& Related Work}
\label{sec:related_work}

\textbf{Automated Code Reviews.}
To help alleviate the cognitive burden of the modern code review process~\cite{compositechanges,workingmemory}, automated code review research introduced three sequential tasks in 1) code change quality estimation~\cite{Li2022CodeReviewer}, 2) review comment generation~\cite{Tufano2022PretrainedModels} and 3) code refinement~\cite{tufano2019learning,thongtanunam2022autotransform,pornprasit2023d}, which mirrors the human process of 1) assessing if a code change is problematic, 2) providing a review comment in natural language that details issues and relevant fixes and 3) addressing the review by revising the code. 
Recent study have leveraged large-scale datasets from open-source platforms to train models to perform these tasks~\cite{tufano2019learning,tufano2021towards,Li2022CodeReviewer}.   
Despite advancements achieved, automating the code review process remains challenging, particularly in the review comment generation task~\cite{tufano2024code}. 


\textbf{Review Comment Generation.}
Review comment generation is formulated as a sequence-to-sequence generation task, where models generate natural language review comments (\rnl) based on code changes (\cdiff). Initially, Tufano et al.~\cite{Tufano2022PretrainedModels} trained a T5 transformer~\cite{raffel2020} on Java (\cdiff, \rnl) pairs, demonstrating that review comments possess latent statistical properties that can be learned. Li et al.~\cite{Li2022CodeReviewer} introduced a large-scale dataset called CodeReviewer which includes nine programming languages for code review-specific pre-training and fine-tuning. Subsequent research using the CodeReviewer dataset explored various solutions, such as joint-training on multiple code review tasks~\cite{sghaier24} and prompting LLMs~\cite{tufano2024code,LLaMA-Reviewer}. Although these models show improvements to the state-of-the-art, their generated comments still fall short when compared to human reviews, raising concerns about their practicality. While these studies focus on developing new techniques, the quality of underlying datasets has been overlooked. This importance of data quality is indeed underscored by a case study at Google~\cite{google2024}, which highlights the potential to build practical code review tools using high-quality data.

\textbf{Review Comment Quality.}
Review comments serve multiple communication intentions, from suggesting code changes to seeking further information or expressing attitudes~\cite{ebert2018communicative}.
Such multi-faceted nature of these comments may hinder automated code review. 
Recent studies~\cite{frommgen2024resolving, tufano2021towards} have highlighted that not all review comments are suitable for training machine learning and artificial intelligence models. This issue is compounded by the fact that datasets mined from open-source platforms often contain noise, such as unclear or non-actionable comments, even after processing with heuristic rules or removing bot-posted comments~\cite{tufano2024code}.
Although the importance of data quality has been recognized, most studies do not focus on identifying and removing noisy review comments at scale, and instead build on them directly. 

\textbf{Existing Approaches for Dataset Quality.} Existing approaches for cleaning code review datasets have primarily relied on manually crafted heuristic rules, which are often dataset-specific and limited in scope. Early work by \cite{gupta2018intelligent} used keywords to develop regular expressions  for C\# to remove non-actionable comments. Similarly, Tufano et al.~\cite{tufano2021towards, Tufano2022PretrainedModels} designed a set of heuristic rules for Java that focused on surface-level features like comment length and keywords (e.g., \textit{less than 5 words and contains `pr'} is considered as noisy). These approaches, however, lack semantic understanding and cross-language generalizability.

More recent efforts have attempted to combine rule-based approaches with supervised machine learning. In developing the CodeReviewer dataset, Li et al.~\cite{Li2022CodeReviewer} applied existing heuristic rules \cite{Tufano2022PretrainedModels} and supplemented them with a machine learning classifier (SVM) trained on manually labeled data.\footnote{Details are provided in their supplementary material (Section A.1 and A.2) at \url{https://arxiv.org/pdf/2203.09095v1}.} Despite these advances, a recent manual analysis ~\cite{tufano2024code} reveals that substantial noise persists in current benchmark datasets including CodeReviewer~\cite{Li2022CodeReviewer}. 
Particularly, the remaining noisy comments require semantic understanding of both code changes and review comments - a capability beyond existing approaches.

\begin{figure*}
    \centering
\includegraphics[height=4.5cm, keepaspectratio]{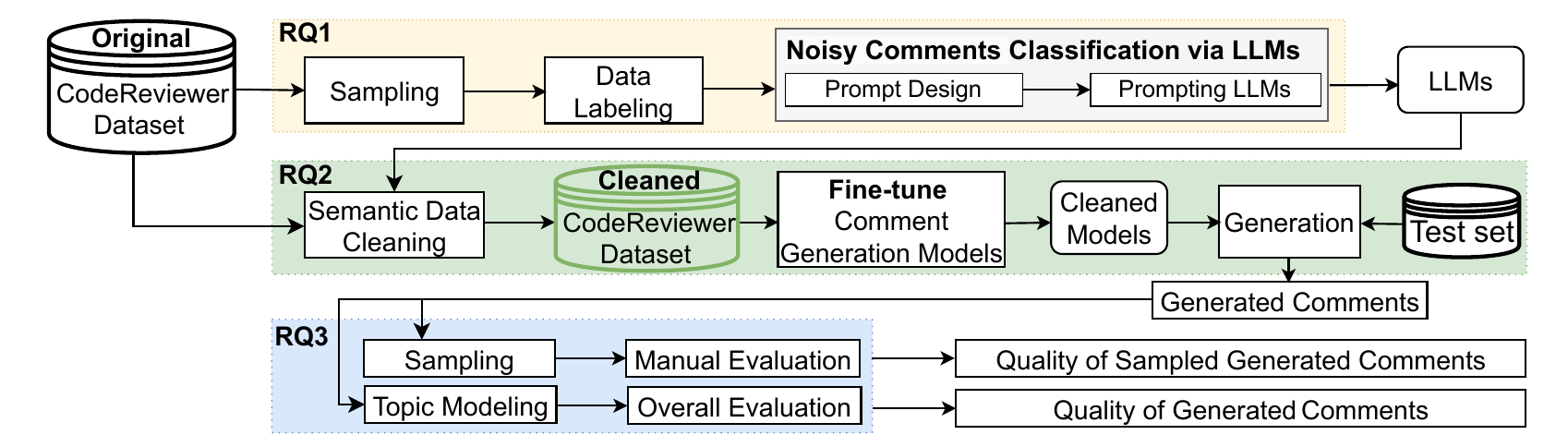} 
    \caption{An overview of the pipeline of our study.}
    \label{fig:framework}
\end{figure*}

\section{Study Design}
\label{sec:study_design}

Our study aims to explore an automated approach that automatically distils human code review comments before using them to train automated comment generation models. 
By focusing on comments that directly contribute to code improvements, we can enhance the quality of training data and, consequently, improve the performance of automated code review comment generation models.

We consider valid and noisy comments using the following definitions, which align with definitions of the prior work~\cite{tufano2024code}.

\begin{itemize}
    \item \textbf{Valid Comments} refer to the review comments that should provide clear suggestions aimed at improving the source code. Given the submitted code change (i.e., code diff), the valid comment should explicitly express the issues, and clearly outline necessary actions to improve the code. The type of requested actions should also be clear, such as refactoring the code to improve code quality (regarding documentation, style, programming conventions and more), writing tests, aligning with good object-oriented design principles, fixing bugs, enhancing logging, or addressing other specific needs.
    
    \item \textbf{Noisy comments} refer to the review comments that do not request direct and applicable actions to refine the code, or the message expressed is unclear and difficult to understand. This includes comments that do not explicitly ask for specific changes, merely justifying the submitted code change, or are of low quality due to vagueness, ambiguity, or other factors that hinder understanding.
\end{itemize}
This definition is consistent with characteristics of useful comments perceived by practitioners~\cite{Bacchelli-etal-2013-Expectations,bosu2015characteristics,turzo}, widely adopted in prior work~\cite{meyers-etal-2018-dataset,Rahman2017Predicting} and corresponds with established taxonomies for comment type classification~\cite{Mantyla-Lassenius-2009-TypesofDefects,tufano2024code}.







\subsection{Research Questions}

The goal of this study is to examine the feasibility of Large Language Models (LLMs) to identify valid and noisy review comments and to investigate the impact of noisy comments on the performance of automated review comment generation models.
To this end, we address three research questions.

\textbf{RQ1: To what degree can large language models semantically clean code review comments?} 
While heuristics have been applied to clean the review datasets~\cite{gupta2018intelligent,tufano2021towards,Tufano2022PretrainedModels}, they have missed noisy comments requiring a deeper understanding of semantics~\cite{tufano2021towards}. Despite the recent advancement of LLMs in various code-related tasks~\cite{zan-etal-2023-large}, their ability to determine the quality of
code review comments has not yet been investigated. Our RQ1 aims to bridge this gap by empirically assessing the effectiveness of LLMs in classifying valid and noisy code review comments, shedding light on an efficient way of improving code review data quality. 

\textbf{RQ2: Does semantic data cleaning impact the accuracy of code review comment generation models?}
Recent work~\cite{tufano2024code} has observed that a substantial proportion (32\%) of the comments remain noisy. 
While retaining only predicted valid comments could improve data quality, it also reduces the training size, potentially affecting model performance. Therefore, we set out RQ2 to examine the impact of data cleaning on the performance of comment generation models.
Specifically, we reassess the accuracy of code review generation models when trained on original versus cleaned datasets.


\textbf{RQ3: Does semantic data cleaning improve the quality of code review comment generation models?} As the quality of training data improved (i.e., a higher ratio of valid comments), the review comment generation models would learn more on valid examples, thus they are more likely to generate high-quality review comments.
Yet, it is unclear to what degree can the quality of generated review comments be improved using cleaned datasets.
Therefore, in RQ3, we aim to examine the improvement in the quality of comments generated from models trained on original and cleaned datasets.
We evaluate the quality in two main aspects, i.e., information and relevance using both manual and semi-automated analyses.

\subsection{Dataset}
In this study, we use the CodeReviewer dataset~\cite{Li2022CodeReviewer} as our subject of study. We selected this dataset because it is the largest real-world and widely used code review dataset, representing typical code review data mined from public platforms like GitHub. This dataset exhibits several key characteristics: (a) it is large in scale, including approximately 116K samples in the training set, along with about 10K samples each in the validation and test sets; (b) it is diverse, containing projects from over 1,000 software repositories and covering nine commonly used programming languages.

\subsection{Study Overview}
Figure~\ref{fig:framework} presents an overview of our study. To investigate the degree to which valid and noisy comments can be automatically classified by LLMs, RQ1 evaluates the accuracy of different prompts and LLMs based on manually annotated samples.  
The most effective approach (prompt strategy and LLM) will be selected to clean the whole training dataset.

RQ2 and RQ3 aim to study the impact of semantic data cleaning on the performance of the comment generation models. We employ the LLM to clean the training data by (a) labeling all instances as valid or noisy, and (b) retaining only instances predicted as valid for a clean dataset. Subsequently, we fine-tune the comment generation models on both the cleaned and original datasets. We then compare their performance in terms of accuracy and quality.
RQ2 evaluates the accuracy of the generated comments by comparing them with actual review comments written by human reviewers.
RQ3 evaluates the quality of the generated comments by examining the information they provide and their relevance to the corresponding code diff. We conduct a manual analysis on a subset of sampled comments and perform a semi-automated analysis to estimate the quality for the overall test set. Below, we describe the details of our experimental design and present experimental results for each RQ (Sections \ref{sec:identifying_noisy_comments_rq1} - \ref{sec:noisy_impact_on_generated_comment_quality_rq3}).



\section{Semantic Data Cleaning via LLMs (RQ1)}
\label{sec:identifying_noisy_comments_rq1}
To assess the effectiveness of large language models (LLMs) in classifying valid and noisy comments, we frame this as a binary classification task. Given a natural language review comment \rnl{} and a code change \cdiff, an LLM classifies whether the comment is valid or noisy based on the definitions described in Section~\ref{sec:study_design}. Due to the lack of an evaluation set, we randomly sampled a subset from the training data and manually categorized it with \tvalid \ and \tnoisy \ labels (Sec~\ref{ssec:rq1:data_labeling}). Using this labeled dataset, we then assessed several LLMs with various prompts to evaluate their performance on comment classification (Sections~\ref{ssec:rq1:experimental_setup} - ~\ref{ssec:rq1:experimental_results}).

\subsection{Data Labeling} 
\label{ssec:rq1:data_labeling}


To address RQ1, we manually labeled a subset of 270 samples that were randomly selected from the training dataset, constituting a statistically significant sample size with a confidence level of 90\% and a margin of error of $\pm$5\%.

The labeling was conducted by two annotators (i.e., the authors of the paper) who have backgrounds in computer science and software engineering with more than five years of programming and software development experience.
In the initial labeling phase, Annotators 1 and 2 independently annotated 50 of the 270 samples based on the definitions of valid and noisy described in Section~\ref{sec:study_design}.
The initial annotation resulted in a Cohen's kappa coefficient of 0.57, indicating moderate agreement.
After discussions on the disagreed cases, the Cohen's kappa coefficient improved to 0.83, signifying near-perfect agreement. The remaining disagreements were resolved by involving Annotator 3 who is a senior researcher with extensive expertise in software engineering.

After the initial phase, the definitions of valid comments were elaborated to become guidelines for determining valid comments based on the shared understanding of the two annotators.\footnote{The complete guideline is available in the replication package~\cite{our_replication_package}.} 
Following this guideline, Annotator 1 proceeded to label the remaining 220 samples, consulting with Annotators 2 and 3 on ambiguous cases to finalize the labels.
Figure~\ref{fig:code_reviews_stacked} provides examples of valid and noisy review comments that we annotated.

Ultimately, out of 270 review comments in the training dataset, the sample comprised 172 valid comments and 98 noisy comments.
The proportion of noisy comments (36\%) is similar to the manual annotation in the test set of CodeReviewer by Tufano et al.~\cite{tufano2024code}, i.e., 32\%,\footnote{{
While Tufano et al. reported a 25\% overall noise ratio across all three benchmarks in their study, our 32\% specifically refers to the CodeReviewer benchmark, which was obtained from their replication package.}} suggesting that our manual annotation is consistent with the prior work.



\subsection{Review Comment Classification}
\label{ssec:rq1:experimental_setup}
\textbf{Large-language Models (LLMs).}
We focused on state-of-the-art LLMs with instruction-following capabilities and exposure to code-related tasks, as this allows the models to follow our classification guidelines and understand the code change.
We selected one commercial model (GPT-3.5) and two open-source models (CodeLlama~\cite{roziere2023code} and Llama 3~\cite{llama3}).
\begin{itemize}
     \item \textbf{GPT-3.5} is an LLM that has been pre-trained on extensive natural language and code corpora. It has been widely used in many code-relevant tasks~\cite{Zheng2024, guo2023exploring}. We employed the gpt-3.5-turbo-0125 in our experiments. 
    \item \textbf{CodeLlama} is a variant of the open-source model Llama2~\cite{touvron2023llama}, tailored for code generation and understanding. We chose the CodeLlama-34b-Instruct version which is fine-tuned to follow human instructions, enabling various tasks across multiple programming languages without task-specific training.

    \item \textbf{Llama3} is a recent large open-source model.
    Despite not being specifically code-trained like CodeLlama,
    Llama3's performance is optimized for coding and reasoning,
    which is suitable for our classification task. We used the Llama-3-8B-Instruct version to enable instruction following in our experiments.
\end{itemize}
{Our model selection considers (a) open-source and closed-source; (b) model size; (c) general vs code-focused. Although CodeLLama and Llama3 are from the same family, they were trained for different purposes, i.e., CodeLLama represents code-focused and Llama3 represents general-purpose LLMs.}



\begin{figure}
    \centering
    \includegraphics[width=0.8\linewidth]{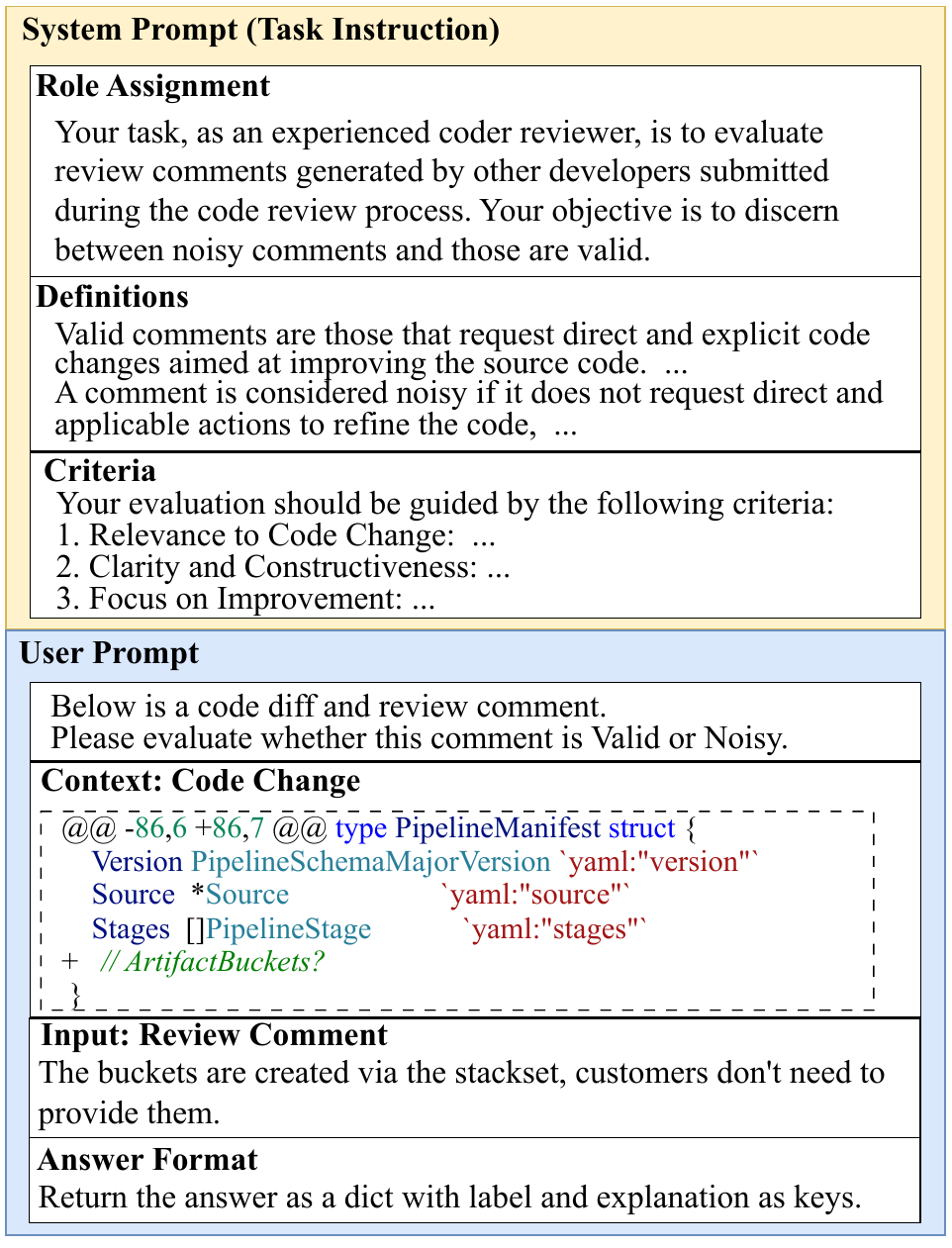}
    \caption{The prompt template for noisy classification using \pdef{} with context.}
    \label{fig:prompt_template}
\end{figure}

\textbf{Prompt Design}. A prompt is used to instruct LLMs to perform specific tasks~\cite{Liu-etal-2023-PromptSurvey}. 
Typically, a prompt comprises two components: the system prompt, which guides the models' overall behaviours, and the user prompt, which provides specific input for each query. Well-crafted and informative prompts in both components can effectively elicit relevant and accurate responses from LLMs. For our review comment classification task, we developed a comprehensive prompt template that draws upon strategies from OpenAI's gpt-best-practices~\cite{openai2023gptbestpractices} and those that have proven effective in software engineering tasks such as software vulnerability detection~\cite{Zhang-etal-2024-PromptEnhanced} and security code review~\cite{yu2024securitycodereviewlarge}. Our system prompt incorporates task instruction details including role assignment, key definitions, and classification criteria. 
The user prompt provides the context of the code change, the specific review comment to be classified, and a structured format for the model's response. Figure~\ref{fig:prompt_template} provides an example of our prompt template. 

To investigate the effectiveness of different components of the prompts on the review comment classification task, we designed four prompts by varying the task instruction in the system prompt and the user prompt. For task instructions, we varied between (a) \pdef: a prompt that includes the definitions of valid and noisy comments as we used in manual annotation, and {(b) \paux: a prompt that supplements the \pdef{} with seven auxiliary rules that concretize the same criteria in Fig~\ref{fig:prompt_template}. These auxiliary rules were developed 
based on our discussions and shared understanding during our initial labelling phase, and were used consistently to guide annotations throughout both RQ1 and RQ2. Complete prompts for \pdef{} and \paux{} for our experiments are in our replication package~\cite{our_replication_package}.}
For the user prompt, we varied between (a) providing only a review comment \rnl{} as input, and (b) providing both the code diff patch \cdiff{} and the corresponding comment \rnl{} together as input. We experimented with these prompts on three LLMs, resulting in 12 experiments. For all models, we set the temperature to a low value of 0.1 to ensure consistency.

\subsection{Evaluation}
To assess the performance of LLMs in classifying valid and noisy comments, we employ \textbf{precision}, \textbf{recall}, and \textbf{F1} metrics. 
To provide a comprehensive view of classification performance, we report class-wise metrics, treating both valid and noisy as positive labels in separate evaluations. This evaluation offers insights into each LLM's ability to identify valid and noisy comments.
We also measure an overview performance considering both valid and noisy classes.
Given the imbalanced nature of our dataset between valid and noisy classes, we report weighted overall performance metrics, where weights are proportional to the sample size of each class.

\input{tables/RQ1_results}

\textbf{Baseline.}
We use the cleaning approach of Li et al.\cite{Li2022CodeReviewer} as a baseline.
When constructing the CodeReviewer dataset, they applied extensive cleaning methods using both heuristic rules~\cite{Tufano2022PretrainedModels} and SVM classifiers.\footnote{The data cleaning approach described in their appendix: \url{https://arxiv.org/pdf/2203.09095v1}}
Consequently, all instances in the CodeReviewer dataset were considered valid according to the approach of Li et al., which implies that any sampled instances are also considered valid by their approach. This baseline is appropriate because state-of-the-art cleaning approaches were already applied; thus,  the remaining noisy ones represent cases that prior techniques could not detect, reflecting the upper bound performance of prior approaches. Other techniques~\cite{tufano2021towards,bosu2015characteristics} are not suitable as they either require additional information that is not available in the CodeReviewer dataset~\cite{bosu2015characteristics}, or are limited to the specific context~\cite{tufano2021towards}.
For example, we applied the heuristic rules of Tufano et al.~\citep{tufano2021towards}, and none of the comments in the CodeReviewer dataset are identified as noisy.



\subsection{Experimental Results}
\label{ssec:rq1:experimental_results}
Table~\ref{tab:rq1_main_results} presents our experimental results on the noisy comments classification task. 
The precision of the baseline on \tvalid{} comments suggests that the original dataset has a valid comment ratio of 63.7\%.

Table \ref{tab:rq1_main_results} shows that
 LLMs achieve an overall F1 of up to 71.7\%. Among the studied LLMs, Llama3 and GPT-3.5 achieved comparable overall precision (71.8\% and 70.3\% respectively), with Llama3 and CodeLlama showing similar performance in recall (72.6\% and 65.6\%). For identifying valid comments, GPT-3.5 and Llama3 achieve a precision of 85.1\% and 75.3\% respectively. This suggests that by retaining only instances predicted as valid, the valid comment ratio is improved by 21.4 and 11.6  percentage points from the original dataset (63.7\%). These findings suggest that LLMs can  distinguish between valid and noisy code review comments.



{Table~\ref{tab:rq1_main_results} shows that the {} prompt with only the review comment (\rnl) as input generally performed best across valid and noisy classes. It achieves the highest precision on overall and \tvalid, and the second highest precision for \tnoisy. Specifically, for Llama3, the performance remains similar whether using \pdef{} or \paux{} prompt. 
For CodeLlama, we observe a performance improvement when only the review comment (\rnl) is provided with the \paux{} prompt. 
Similarly, providing additional code context (\cdiff) alongside review comments (\rnl) does not improve the model performance. 
For instance, CodeLlama has a precision drop of 9.9\% - 24\% when \cdiff{} is provided. 
This may be because the additional details of \paux{} and code context increase the prompt length and distract the models, potentially leading them to overlook crucial criteria in longer prompts~\cite{liu2024lost}. For instance, we observed that GPT-3.5 returns unexpected labels or empty responses for three samples with long context.
These findings highlight the importance of concise, focused inputs for optimal LLM performance.
\begin{tcolorbox}[size=title]
{\textbf{Answer to RQ1:}} LLMs show promising potential in classifying code review comments, with an overall F1 up to 71.7\%. The precision in identifying valid comments highlights that the proportion of valid comments can be improved from 64\% on the original dataset to 85\% on our cleaned datasets.
\end{tcolorbox}

\section{Impact on Comment Generation Accuracy (RQ2)}
\label{sec:noise_impact_on_comment_generation_models_rq2}
To investigate the impact of semantic data cleaningon comment generation models, we employ the LLMs from RQ1 to clean the dataset. Then, we fine-tune the pre-trained models with the cleaned dataset and evaluate the performance of the models in generating review comments.

\subsection{Semantic Data Cleaning}
To clean the dataset, we use the LLMs to predict the \tvalid\ and \tnoisy\ classes and retain the \tvalid\ instances (i.e., removing the \tnoisy\ instances). We used (1) \textbf{GPT-3.5} and (2) \textbf{Llama3} using the \pdef{} prompt with \rnl\ since these two models exhibit strong performance in retaining a high ratio of \tvalid\ comments (with 85.1\% and 75.3\% precision, respectively) while exhibiting complementary characteristics in recall. This allows us to obtain cleaned datasets with varying degrees of a valid comment ratio and training size.

We use the LLMs to clean the training and validation sets of the studied dataset. Table~\ref{tab:dataset_statistics} shows the statistics summary of cleaned training and validation set sizes with different LLMs. The \textsc{Original} row shows the size of the original training and validation set of CodeReviewer~\cite{Li2022CodeReviewer}.
The \textsc{Cleaned} rows show the size of the training and validation cleaned by \textbf{GPT-3.5} and  \textbf{Llama3} using the \pdef{} prompt with \rnl{}.
The \textsc{Controlled} are the training and validation sets that are randomly sampled to have the same number of instances as the two cleaned datasets.
These control groups are designed to account for the impact of reduced data in clean datasets on performance, as the removal of noisy data results in fewer instances for training and validation.


\begin{table}[!t]
    \centering
    \caption{Statistics of Datasets for Comment Generation.}
    \begin{tabular}{@{}lcc|c@{}}
        \toprule
        \textbf{Dataset} & \textbf{Training set} & \textbf{Validation set} & \textbf{Test set} \\
        \midrule
        \textsc{Original} & 117,739 & 10,319 & \multirow{5}{*}{10,169} \\ \cmidrule{1-3}
        \cleangpt & 39,625 & 3,395 & ~\\
        \cleanllama & 87,872 & 7,571 &  ~\\ \cmidrule{1-3} 
        \controlgpt & 39,625 & 3,395 &  ~\\
        \controlllama & 87,872 & 7,571 & ~ \\
        \bottomrule
    \end{tabular}
    \label{tab:dataset_statistics}
\end{table}
\subsection{Comment Generation Models}

To address RQ2, we focused on models that can generate code review comments. Our study requires code review models with reproducibility for a fair comparison of model performance between original and cleaned datasets under the same settings. Specifically, we need models that provide publicly available checkpoints and fine-tuning scripts. While prior work has fine-tuned various code models~\cite{auger,CCT5,sghaier24,Li2022CodeReviewer} and LLMs~\cite{LLaMA-Reviewer} for this task, many do not meet our criteria. Therefore, we selected two widely known models for code review automation, i.e., CodeT5~\cite{wang-etal-2021-codet5} and CodeReviewer \cite{Li2022CodeReviewer} that met our criteria.

\textbf{CodeT5} is a general-purpose encoder-decoder Transformer model pre-trained on both programming and natural languages, which demonstrated effectiveness across multiple downstream tasks. 
\textbf{CodeReviewer} is a state-of-the-art model pre-trained on tasks relevant to code changes. It leverages the pre-trained weights from CodeT5 for model initialization and continues pre-training on datasets pertinent to code reviews.
This specialized pre-training allows CodeReviewer to demonstrate superior performance on the comment generation task. 

{We select CodeT5 and CodeReviewer as representative models because we aim to compare the impact of data quality between a general code-pretrained model (CodeT5) and a code review-specific pretrained model (CodeReviewer). This comparison allows us to understand whether the benefits of data cleaning generalize across different models and training data.
We select CodeT5 over alternatives such as CodeBERT because CodeT5 has consistently demonstrated superior performance across various code-related tasks~\cite{wang-etal-2021-codet5}. 
For CodeReviewer, recent studies have validated its continued competitiveness in the code review domain. 
For instance, a recent study by Google researchers~\cite{frommgen2024resolving} highlighted CodeReviewer as ``perhaps the closest recent result" to their code review assistant trained on their high-quality industrial datasets. Furthermore, Fan et al.~\cite{fan2024exploring} found that fine-tuning more recent large language models like Llama2 and CodeLlama sometimes did not outperform CodeReviewer, with BLEU-4 score differences ranging from -1.28 to 0.42. These findings demonstrate that CodeReviewer remains a competitive baseline for code review automation, making it an appropriate choice for evaluating the impact of data quality on model performance.}

\textbf{Fine-tuning}: We fine-tuned the CodeReviewer and CodeT5 models using original, cleaned, and controlled datasets. All experiments were conducted on four NVIDIA H100-80GB GPUs.We followed the hyperparameters specified in the original CodeReviewer, with one exception: we adjusted the batch size from 64 to 32, which resulted in an improved BLEU score from 5.3 to 5.7 and improved the training efficiency. To avoid over-fitting, we used an early stopping criterion that ended training after 5 epochs without improvement on validation set.

    


\begin{table*}[h]
        \centering
        \caption{Model Performance (BLEU-4) on Comment Generation Models.}
        \label{table:comment_gen_main_results}
        \setlength{\tabcolsep}{5pt}  
        \begin{tabular}{ll|l|ll|ll|ll}
        \toprule 
        \textbf{$M$} & \textbf{Dataset} & \textbf{Test} & \textbf{Valid$_{\text{Our\&Tufano}}$} & \textbf{Noisy$_{\text{Our\&Tufano}}$} & \textbf{Valid$_{\text{Our}}$} & \textbf{Noisy$_{\text{Our}}$} & \textbf{Valid$_{\text{Tufano}}$} & \textbf{Noisy$_{\text{Tufano}}$} \\ \midrule
        \multirow{5}{*}{\rotatebox{90}{CodeReviewer}} & \textsc{Original} & 5.73 & 6.17 & 5.41 & 5.45 & 5.17 & 7.12 & 5.60  \\
        ~ & \cleangpt &  \textbf{6.04} \ 5.4\%↑$^{*}$ & \textbf{6.97} \ \textbf{13.0\%}↑$^{*}$  & 5.02 \ 7.2\%↓ & \textbf{5.93} \ 8.8\%↑ & 5.19 \ 0.4\%↑ & \textbf{7.99} \ \textbf{12.2\%}↑$^{*}$ & 4.83 \ 13.8\%↓ \\
        ~ & \controlgpt & 5.63 & 6.20 & 5.43 & 5.21 & 5.13 & 7.39 & 5.70 \\ 
        &  \cleanllama & \underline{5.97} \ 4.2\%↑$^{*}$ & \underline{6.63} \ 7.5\%↑$^{*}$  & 5.18 \ 4.3\%↓ & 5.64 \ 3.5\%↑& 5.11 \ 1.2\%↓ & \underline{7.71} \ \underline{8.3\%}↑$^{*}$ & 5.14  \ 8.2\%↓ \\
         ~ & \controlllama & 5.63 & 6.18 & 5.66 & 5.12 & 5.36 & 7.45 & 5.86 \\
        \midrule 
        \multirow{5}{*}{\rotatebox{90}{CodeT5}} & \textsc{Original}&  5.19 & 5.34 & 5.04 & 4.84 & 5.09 & 5.85 & 6.03  \\ 
        ~ & \cleangpt & 5.67 \ \textbf{9.2\%}↑$^{*}$ & 6.00 \ \underline{12.4\%}↑$^{*}$  & 5.23 \ 3.8\%↑$^{*}$  & \underline{5.88} \ \textbf{21.5\%}↑$^{*}$ & 5.27 \ 3.5\%↑ & 6.06 \ 3.6\%↑ & 5.15 \ 14.6\%↓ \\
        ~ & \controlgpt & 5.20 & 5.34 & 5.30 & 5.17 & 5.39 & 5.45 & 5.41  \\ 
         ~ & \cleanllama& 5.54 \ \underline{6.7\%}↑$^{*}$ & 5.74 \ 7.5\%↑$^{*}$  & 5.33 \ 5.8\%↑ & 5.32 \ \underline{9.9\%}↑$^{*}$ & 5.14 \ 1.0\%↑ & 6.09 \ 4.1\%↑ & 5.46 \ 9.5\%↓ \\
        ~ & \controlllama& 5.21 & 5.19 & 5.12 & 4.95 & 5.26 & 5.38 & 5.01 \\
       
        \bottomrule 
        \multicolumn{9}{l}{\footnotesize The highest and second-highest results are in bold and underlined, respectively. $^{*}$ indicates the statistical significance (p-value $<$ 0.05).
        } \\
        \end{tabular}
        \end{table*}

\subsection{Evaluation} 
We evaluate the fine-tuned models using the original test set.
We do not clean the test set because our goal is to assess the impact of different training sets on comment generation. 
Nevertheless, there may be noisy comments in the test set.
Thus, we evaluate models' performance on valid and noisy samples. To be consistent with RQ1, we manually labeled a subset of review comments from the test set as valid or noisy using our guidelines described in Section~\ref{ssec:rq1:data_labeling}. We randomly sampled 371 review comments, ensuring a significant sample size with a confidence level of 95\% and a margin of error of $\pm$5\%. The annotation was conducted by Annotator 1, with ambiguous cases discussed with Annotator 2 to reach a consensus. As a result, this sample includes 223 valid samples and 148 noisy samples. To further increase the generalisation of the results, we obtained a manually labeled subset sampled from the test set by Tufano et al~\cite{tufano2024code}, which includes 234 valid comments and 135 noisy comments that were randomly sampled from the same CodeReviewer test set that we used. We combined the two labeled test sets, resulting in a total of 726 samples, comprising 452 valid and 274 noisy comments.

To evaluate the quality of generated comments, we conduct automatic evaluation using the BLEU (Bilingual Evaluation Understudy) metric~\cite{papineni-etal-2002-bleu}, which quantifies the n-gram lexical overlap between the generated comments and the ground truth comments from human reviewers. Following prior work~\cite{Li2022CodeReviewer}, we use the BLEU-4 variant to calculate the overlap of up to 4-grams between generated and ground truth comments. We evaluate model performance on the entire test set and on manually labeled valid and noisy subsets. We use the one-sided Wilcoxon signed-rank test to statistically examine the difference in BLEU-4 between the original and cleaned models.


\subsection{Experimental Results}
Table~\ref{table:comment_gen_main_results} presents our experimental results. Overall, despite the substantial reduction in training data (i.e., 66\% smaller using GPT-3.5 and 25\% smaller using LLAMA3), the performance of comment generation models is not negatively impacted. Instead, when training using \textsc{Cleaned} datasets, the BLEU-4 increases by 4.2\%-5.4\% for CodeReviewer models and 6.7\%-9.2\% for CodeT5 models compared to the \textsc{Original} dataset. The Wilcoxon signed rank tests also confirm that BLEU-4 of the cleaned models is statistically higher than the original models for all test instances. The improvement aligns with previous work; e.g., $\Delta$BLEU(CodeReviewer, CodeT5) = 0.49~\cite{Li2022CodeReviewer}, and fine-tuning LLMs also demonstrates similar gains, e.g., $\Delta$BLEU(Llama-Reviewer, CodeReviewer) = 0.4~\cite{LLaMA-Reviewer}.
However, merely reducing the training data does not improve performance, as evidenced by the results of the \textsc{Controlled} groups. This indicates that training data quality is as important as data quantity. 


Models trained on cleaned datasets demonstrate consistent improvements in the valid subsets.
When considering both our and Tufano's valid subsets (\textbf{Valid$_{\text{Our\&Tufano}}$}), CodeReviewer exhibits substantial gains, with increases ranging from 7.5\% to 13.0\% in BLEU-4 scores. Similarly, CodeT5 achieves increases between 7.5\% and 12.4\%. 
Table~\ref{table:comment_gen_main_results} also shows a consistent increase in BLEU-4 for \textbf{Valid$_{\text{Our}}$} and \textbf{Valid$_{\text{Tufano}}$} independently.
These results strongly indicate that models trained on cleaned data generate comments that more closely align with valid human review comments.
On the other hand, the performance of \textsc{Cleaned} models on the noisy subsets is inconsistent. 
Nonetheless, these BLEU-4 score variations may not accurately reflect the true quality of generated comments due to the inherent noise in the noisy dataset. 
Therefore, we conduct a manual evaluation in RQ3.

\begin{tcolorbox}[size=title]
{\textbf{Answer to RQ2:}}
Despite the training size reduction, the performance of comment generation models is improved when using cleaned datasets.
Specifically, the cleaned models perform consistently better on valid comments in test sets, leading to even higher increases in BLEU-4 scores of 13.0\% - 12.4\% compared to the original models.
\end{tcolorbox}

\section{Impact on Generated Comment Quality (RQ3)}
\label{sec:noisy_impact_on_generated_comment_quality_rq3}
While BLEU-4 evaluates accuracy in terms of lexical correspondence, it does not account for the diversity in how similar intents can be expressed~\cite{Stapleton-etal-2020-humanstudy}. 
In addition, the quality of comments extends beyond lexical similarity to human reviews, incorporating factors such as informativeness and contextual relevance to code changes are crucial for code review~\cite{Oleksii-etal-2016-CodeReviewQuality,Rahman2017Predicting}. This is particularly important given the potential noise in human review comments. Therefore, in RQ3, we evaluate the quality of generated comments.

We assess the quality based on how much information a generated comment provides, and how relevant the comment is to the code change \cdiff{} (Sec~\ref{ssec:comments_quality_aspects}). We compare the generations of the models trained by the \textsc{Original} and \textsc{Cleaned} datasets. We focus the CodeReviewer models in this RQ given their superior performance over the CodeT5 models on comment generation. Our analysis focuses on comments generated by three model variants, i.e., trained by the \textsc{Original} dataset and two cleaned versions —\cleangpt{} and \cleanllama{} datasets — to evaluate comment quality.

We conduct a two-fold evaluation: (a) a manual evaluation of a sampled subset of  generated comments (Sec~\ref{ssec:manual_quality_evaluation}); and (b) an overall quality evaluation of the entire test set (Sec~\ref{ssec:overall_quality_estimation}).

\subsection{Comments Quality Measures} 
\label{ssec:comments_quality_aspects}
Following the definitions used in the CodeReviewer paper~\cite{Li2022CodeReviewer}, we evaluate the information and relevance of generated comments as follows.

For \textbf{information}, we evaluate how informative the comment is for the code author to revise the code change. 
Each comment will be labeled with an information score of one to five, where five indicates very informative.
For example, comments explicitly point out the issues and provide concrete suggestions (e.g., \textit{``Shouldn't this be an assert instead of a throw?''}) will have a higher information score than those purely seeking clarification (e.g., \textit{``Why do we need to change this?''}).  

For \textbf{relevance}, we evaluate to what extent the review comment is related to the corresponding code change. Each comment will be labeled with a relevance score of one to three, where three indicates high relevance. Comments that explicitly point out the location of issues in the code changes will receive high relevance scores, while comments that implicitly indicate the issue location or are not related to the code changes will receive low relevance scores. 
Note that we did not evaluate the logical correctness of comments in our relevance metric. This aspect often requires context beyond the code change and lacks a definitive ground truth which is prone to uncertainty. Thus, we focus on objective factors such as the explicitness of issue location within the code change.

\subsection{Manual Evaluation} 
\label{ssec:manual_quality_evaluation}

We conducted a manual evaluation on a sampled subset of the generated comments.
We randomly sampled 100 instances from our entire labeled test set.
Each instance received three comments generated by the three model variants (\textsc{Original}, \cleangpt, and \cleanllama{}). Thus, we evaluated a total of 300 generated comments. 
To ensure a reliable assessment of comment quality across our sample, the evaluations were manually conducted in two rounds.  
Initially, Annotators 1 and 2 independently evaluated 50 generated comments for both information and relevance according to the guidelines with definitions for each score.\footnote{See information and relevance definitions in our replication package~\cite{our_replication_package}.}
The first round of annotation achieved a Cohen's kappa of 0.71 (substantial agreement) for information and 0.42 (moderate agreement) for relevance. 
Following a discussion to resolve disagreements, a second round of independent annotation of another 50 comments was conducted for relevance, improving the Cohen's kappa to 0.60 (substantial agreement). 
Annotators 1 and 2 annotated the remaining 200 comments independently and then discussed to resolve remaining disagreements. 

\begin{figure*}[t]
    \centering
    \begin{subfigure}[b]{0.24\textwidth}
        \centering
        \includegraphics[width=\textwidth]{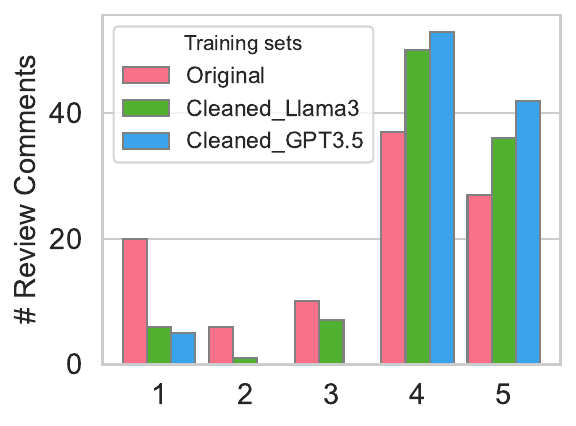}
        \caption{Information score (manual)}
        \label{fig:information_score_distribution_overall}
    \end{subfigure}
    \hfill
    \begin{subfigure}[b]{0.24\textwidth}
        \centering
        \includegraphics[width=\textwidth]{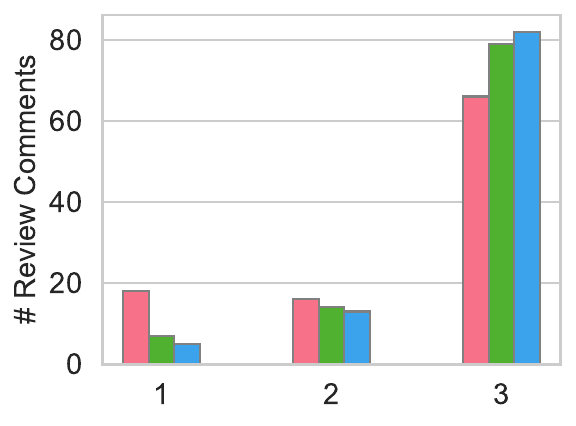}
        \caption{Relevance Score (manual)}
        \label{fig:relevance_score_distribution_overall}
    \end{subfigure}
    \hfill
    \begin{subfigure}[b]{0.24\textwidth}
        \centering
        \includegraphics[width=\textwidth]{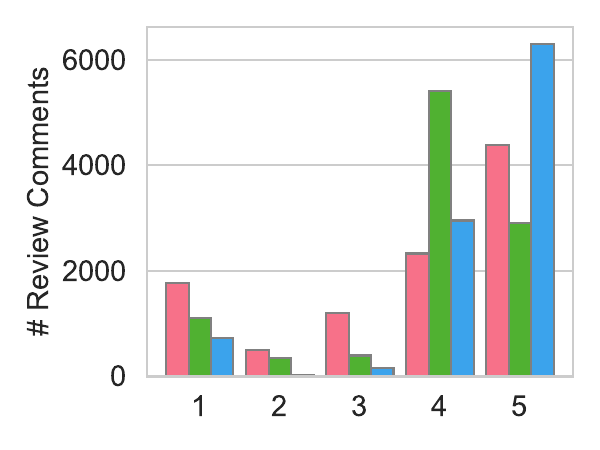}
        \caption{Information Score (overall)}
        \label{fig:topic_information_score_distribtuion}
    \end{subfigure}
     \hfill
    \begin{subfigure}[b]{0.24\textwidth}
        \centering
        \includegraphics[width=\textwidth]{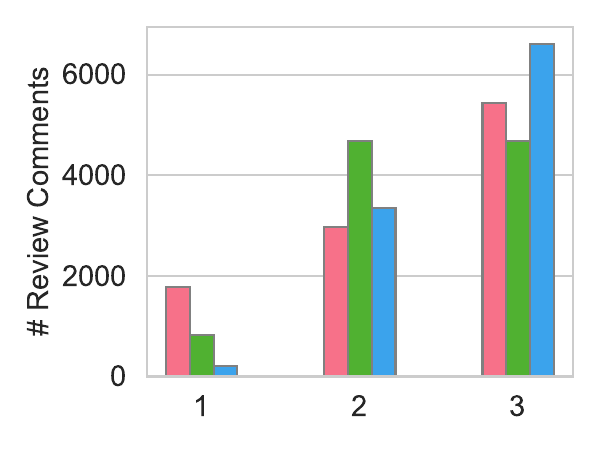}
        \caption{Relevance Score (overall)}
        \label{fig:topic_information_score_distribtuion_relevance}
    \end{subfigure}
    \caption{Distribution of information and relevance scores on tests across CodeReviewer models trained on different training sets.}
    \label{fig:information_score_distributions}
\end{figure*}
\input{tables/information_relevance_scores}

\subsection{Overall Evaluation}
\label{ssec:overall_quality_estimation}

In addition to manual evaluation which is feasible only for a limited number of samples due to its labor-consuming nature, we conducted a semi-automated method to estimate the information and relevance of generated comments for the entire test setTo do so, we employed topic modeling to cluster comments generated by each model,
then we manually annotated the information and relevance scores for each cluster.

\textbf{Topic Modeling.} We employed BERTopic~\cite{grootendorst2022bertopic}, a widely adopted topic modeling technique that outperforms traditional methods like LDA, to extract meaningful clusters from the generated comments. BERTopic is an embedding-based approach that leverages a transformer-based model to represent each comment as a contextual embedding and applies clustering to these embeddings. This method effectively captures semantic similarities in comments and group similar comments in clusters. 
To generate embeddings, we used the recent code model, CodeT5+~\cite{wang-etal-2023-codet5+}, which performs effective bi-modal representation tasks involving both code and natural language. For the clustering model, we used agglomeration hierarchical clustering, which assigns each comment into its own cluster and iteratively merges the closest pairs of clusters until a stopping criterion is met.

We measured the cluster quality using the mean coherence score~\cite{roder-etal-2015-topic-coherence}, which measures how semantically similar the comments within a topic are to each other. 
We set the number of clusters to 50, which achieved coherence scores above 0.67 for all three sets of comment clusters. The high coherence indicates that the clusters are generally well-formed and internally consistent, indicating cohesive clusters for manual analysis.

\textbf{Annotating Comment Quality.} 
We evaluated the quality of generated comments as follows. For each cluster, BERTopic identified the top three representative comments that are most semantically similar to its cluster representation, using cTF-IDF and cosine similarity \citep{roder-etal-2015-topic-coherence}. 
Then, we annotated the information and relevance scores following our established guidelines (Sec \ref{ssec:comments_quality_aspects}). Similar to the manual evaluation, Annotators 1 and 2 independently conducted the evaluations and resolved disagreements through discussion. 
Given the relatively high coherence scores of the clusters, we considered the average information and relevance scores of the representative comments as \textit{approximated} quality scores for all comments in the corresponding cluster.


\begin{figure}
    \centering
    \includegraphics[width=\linewidth]{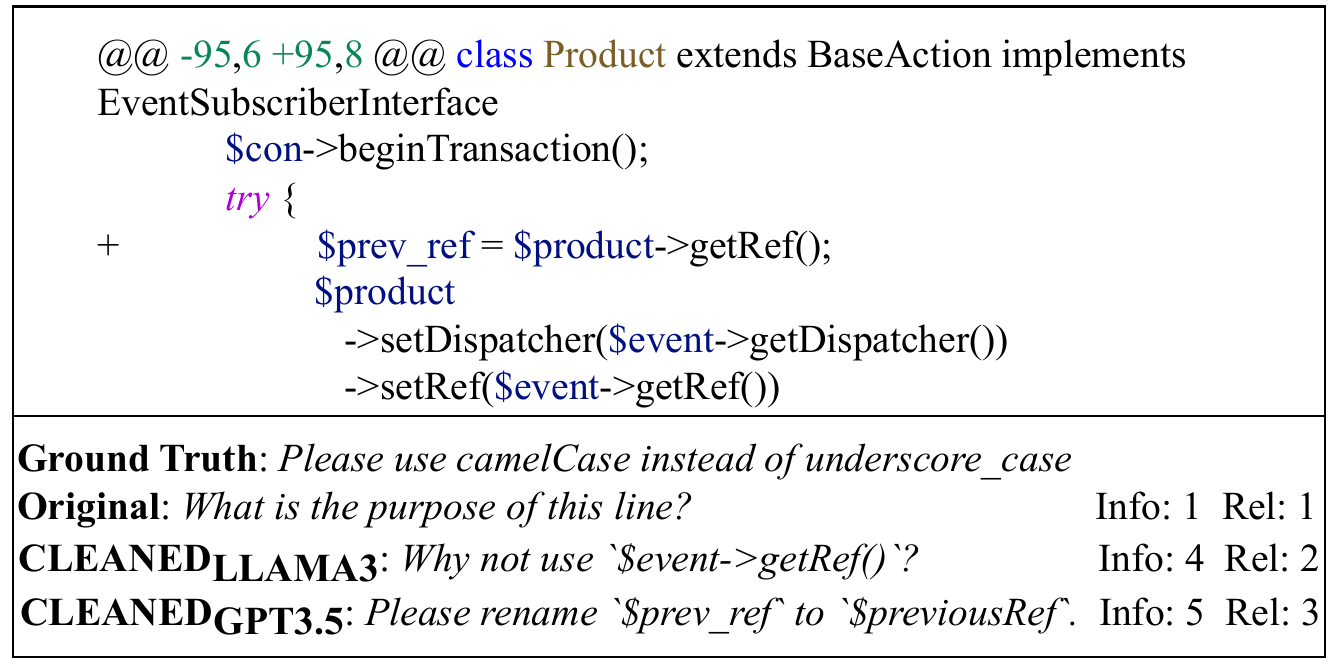}
    \caption{Example comments generated by original and cleaned models with information (Info) and Relevance (Rel) scores.}
    \label{fig:example_generated_comments}
\end{figure}

\subsection{Experimental Results}
Table~\ref{tab:information_relevance_scores} provides the average information and relevance scores. Figure~\ref{fig:information_score_distributions} illustrates the distribution of these scores.

Based on the manual evaluation of 100 sampled instances, the \textsc{Cleaned} models achieve substantial improvement over the \textsc{Original} model. 
For example, \cleangpt{} achieves a 24\% increase in information score and an 11\% increase in relevance score. 
Figure~\ref{fig:information_score_distributions}~(a) and (b) show a clear trend towards higher information and higher relevance for the \textsc{Cleaned} models compared to the \textsc{Original} model. 
The most notable change is a 73-80\% decrease in low information (scores 1 and 2) and a 61-72\% decrease in low relevance (score 1) for both cleaned datasets.
These results highlight a substantial improvement in the quality of comments after cleaning.

We observe that the cleaned models tend to generate comments including code tokens related to the code change, making comments more specific and relevant to the code under review.
The examples in Figure~\ref{fig:example_generated_comments} show that a comment from the \cleangpt{} model points out an issue more directly relevant to a changed code, compared to the comment from the \textsc{Original} model.
The improved relevance likely benefits from the characteristics of valid comments (i.e., more actionable and context-aware) in the training data. 


The results of an overall evaluation for the whole test set align with the manual evaluation results, as shown in Table~\ref{tab:information_relevance_scores}.
For example, the approximated information and relevance scores increase by 18\% and 11\% for \cleangpt{}.
Similar to our manual evaluation, we observe a 35-66\% decrease in low information and a 55-88\% decrease in low relevance for both \textsc{Cleaned} models.
The results suggest that the cleaned datasets improve the quality of generated comments in terms of information and relevance.



\begin{tcolorbox}[size=title]
{\textbf{Answer to RQ3:}}The quality of generated review comments can be improved when using our cleaned datasets. 
Our manual evaluation shows an increase up to 24\% in informativeness scores and 11\% in relevance scores. 
\end{tcolorbox}

\section{Discussions}
\label{sec:discussion}
In this section, we discuss the benefits, limitations, and costs of using LLMs to clean the review dataset.

\textbf{LLMs' Capability:} \textit{To what extent LLMs can classify valid and noisy code review comments? } 
Our RQ1 has shown promising results on leveraging LLMs to classify valid and noisy comments, paving a crucial step towards automating dataset cleaning. Nonetheless, LLMs sometimes struggle to identify noisy comments. We observe that LLMs often incorrectly classify comments including domain-specific terms but do not provide improvement suggestions as valid.
For example, 
\textit{``Why `preexec\_fn' is not set in the previous version?'' includes `preexec\_fn'.} 
This may be because LLMs tend to preserve their learned knowledge (i.e., code tokens), consequently failing to adhere to the classification instruction that valid comments must explicitly address code changes.
This underscores the complexity of noisy review comments that future research can address to improve the performance of noise removal in code review datasets. {Since our current study examines models individually, future research could explore ensemble approaches that leverage common predictions across multiple models to enhance classification accuracy.}


\textbf{Cost-Performance Trade-off}: \textit{What are the costs of removing noisy comments using LLMs, and what benefits can it bring?} We evaluated the efficiency of our LLM-based approach in terms of both time and cost, as well as its impact on the quality of comment generation. 
In terms of costs, GPT-3.5 required \$50 USD and 39 hours to clean the entire dataset and the open-source Llama3 took 15 hours. This cost is lower than manual annotation, which would cost \$25,600 USD based on crowdsourcing rates (\$8/hour)\footnote{\url{https://www.prolific.com/calculator}} assuming 2,000 man-hours (one minute/comment) to annotate the entire training and validation sets.

Given the cost of LLMs, the benefits of cleaned datasets are substantial.
RQ2 shows that the cleaned models achieve an 13\% and 12.4\% increase in BLEU-4 for valid comments and RQ3 shows that the quality of the generated comments substantially increases.
Moreover, we observe that the general-purpose CodeT5 model with a cleaned dataset achieved comparable performance (BLEU-4 of 5.67) to the original CodeReviewer (BLEU-4 of 5.73) while using far fewer resources. CodeReviewer is a code-review specific model which was further pre-trained on CodeT5 with 463.2GB of code review data with 2,481k comments over 250k steps. In contrast, the CodeT5 model trained on \cleangpt{} used only 39k comments (98.4\% fewer) and 7k training steps (97.2\% fewer). 
This highlights the benefits of high-quality data for model efficiency, potentially reducing computational costs and environmental impact in large-scale training.

\section{Threats to Validity}
\label{sec:threats_to_validity}


\textbf{Construct Validity.}
{We define `valid' comments as those that are non-noisy (i.e., not vague, difficult to understand, or seeking clarification). It is possible that these valid comments can be technically incorrect or considered not useful by practitioners. However, assessing such technical correctness requires project-specific expertise to validate the comments.} The classification performance of LLMs may vary with different prompts and hyperparameter settings. 
Different prompt strategies and LLMs might yield different results.
However, our primary goal was not to find the best LLMs nor optimize their hyperparameter settings, but to investigate the feasibility of automatically cleaning the review data using LLMs.
In addition, we only evaluate the quality of generated comments in terms of informativeness and relevance. 
There might be a risk of incorrectness, where models fabricated non-existent code tokens by combining existing ones in the code change.
However, evaluating correctness and hallucination is a non-trivial and manual-intensive task that requires a deep understanding of the system and code change.  



\textbf{Internal Validity.} The manual labeling and evaluation are subject to cognitive biases.
To mitigate these, the annotations were conducted independently, and inter-rater agreements were measured.
In addition, the results were reviewed blindly without knowing which models generated the comments to ensure that the quality scores do not purposefully favor any particular models.
{While the annotation of test samples by us and Tufano et al. could potentially be different, merging these two samples is reasonable as (1) the noisy/valid definitions and the criteria are derived from both Tufano's and other works, and (2) we evaluate models separately on each labeled test set and on the combined set.}
We use topic modeling to cluster the generated comments, and then assign quality scores based on subsets of each cluster. 
It is possible that the scores are inaccurate.
Nevertheless, this overall evaluation is consistent with the manual evaluation of the samples.
It is important to note that this evaluation is just an approximation to complement our manual evaluation. {LLMs used for classification in this work were trained on GitHub data. Thus, they may be susceptible to data leakage. 
Nevertheless, they were not specifically trained to classify noisy code review comments. Therefore, we believe the impact of potential data leakage on our classification task is minimal.}

\textbf{External Validity}
Our study is based on two widely-known models (i.e., CodeT5 and CodeReviewer). 
The findings may not generalize to other code review models and datasets.
{Unfortunately, other existing code review models are not suitable for our study for various reasons, for example, some models~\cite{CCT5,sghaier24, LLaMA-Reviewer} cannot be replicated due to missing model checkpoints or fine-tuning scripts, while ~\cite{auger} was pretrained only with Java examples, which is not comparable with CodeReviewer's multi-language dataset.} {While we acknowledge that newer models continue to emerge, our primary objective was to demonstrate the fundamental impact of data quality on code review generation, rather than achieving state-of-the-art performance. The consistent improvements we observed across both models suggest that our data cleaning approach offers benefits that likely extend beyond a specific model. Furthermore, our study on identifying high-quality datasets remains valuable for emerging LLM-based code review systems in several key ways: (1) enhancing retrieval-augmented generation by validating the quality of retrieved examples, and (2) improving benchmark dataset quality by providing more reliable test sets for model evaluation.}
Therefore, we believe our key finding — high-quality datasets improve comment generation models — remains valid and applicable to a broad range of code review contexts.

\section{Conclusion}
\label{sec:conclusion}
In this paper, we address the critical issue of data quality in code review automation. 
We explore a novel method leveraging large language models (LLMs) to identify and remove noisy comments.
Our results show that LLMs can achieve 66-85\% precision in identifying valid comments, improving the proportion of valid comments from 64\% in the original dataset to up to 85\% in our cleaned datasets. By training code review models on cleaned datasets, we observe substantial improvements in review comment generation quality, with up to a 13\% increase in BLEU-4 scores and a 24\% improvement in informativeness. 
Our work demonstrates the feasibility of automatically cleaning review datasets and offers insights into how data quality affects model performance in automated code review. Future research could investigate advanced cleansing techniques for complex comments.

\section{Acknowledgement}
This research was supported by The University of Melbourne’s Research Computing Services and the Petascale Campus Initiative. Patanamon Thongtanunam was supported by the Australian Research Council’s Discovery Early Career Researcher Award (DECRA) funding scheme (DE210101091).


\bibliographystyle{IEEEtran}
\bibliography{reference.bib}

\end{document}

%% file: tables/example_reviewer_comments_valid_noisy.tex
\begin{figure}[]
\begin{minipage}{0.49\textwidth}

    \includegraphics[width=\linewidth]{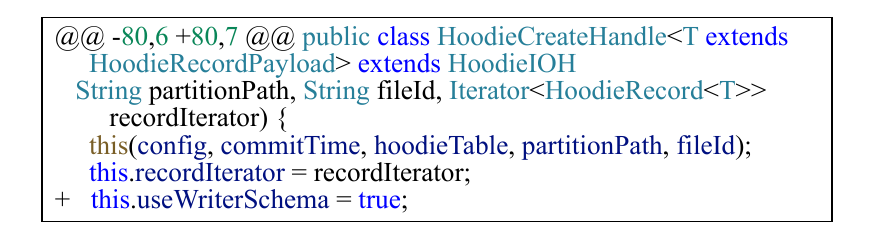}

\begin{small}
\textbf{Reviewer's Comment:} \textit{Why do we have this flag?} \\
\textbf{Label:} Noisy Comment 
 
\end{small}
\end{minipage}

\begin{minipage}{0.49\textwidth}
 \includegraphics[width=\linewidth]{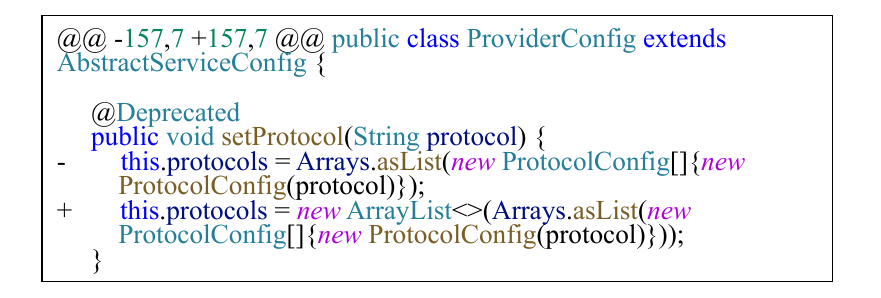}
\begin{small}
\textbf{Reviewer's Comment:} \textit{This can be simplified as  \lstinline|new ArrayList<>(Arrays.asList(new ProtocolConfig(protocol)))|} \\
\textbf{Label:} Valid Comment
\end{small}
\end{minipage}

\caption{Examples of noisy (Top)  and valid (Bottom) comments for automated review comment generation.}
\label{fig:code_reviews_stacked}
\end{figure}

%% file: tables/RQ1_results.tex
\begin{table*}[!th]
\centering
\caption{Experimental results on noisy classification.}
\label{tab:rq1_main_results}
\begin{tabular}{p{1cm} p{2cm} p{1.5cm} | p{0.5cm} p{0.5cm} p{0.5cm} | p{0.5cm} p{0.5cm} p{0.5cm} p{0.5cm} | p{0.5cm} p{0.5cm} p{0.5cm} p{0.5cm}}
\toprule
\textbf{Prompt} & \textbf{Model} & \textbf{Input}  & \multicolumn{3}{c}{\textbf{Overall (weighted)}} & \multicolumn{4}{|c}{\textbf{Valid (172)}} & \multicolumn{4}{|c}{\textbf{Noisy (98)}} \\
\midrule 
 ~ & ~ & ~ & \textbf{Prec} & \textbf{Rec} & \textbf{F1} &  \textbf{Prec} & \textbf{Rec} & \textbf{F1} & \textbf{\#} & \textbf{Prec} & \textbf{Rec} & \textbf{F1} & \textbf{\#} \\ 
\cmidrule(lr){4-6} \cmidrule(lr){7-10} \cmidrule(lr){11-14} 
~ & \textbf{Baseline~\cite{Li2022CodeReviewer}} & - & 40.6 & 63.7 & 49.6 &  63.7 & 100 & 77.8 & 270 & 0 & 0 & 0 & 0 \\ \midrule 
\multirow{7}{*}{\rotatebox{90}{\pdef}} &  \textbf{GPT-3.5} & \rnl  & 70.3 &   54.1 & 55.7 & \textbf{85.1} & 36.6 & 51.2 & 74 & 44.4 & \textbf{88.8} & \textbf{59.2} & 196 \\
& \textbf{CodeLlama} & \rnl  & 64.1 & 65.6 & 58.0 &  66.0 & \textbf{94.8} & 77.8 & 247 & 60.9 & 14.3 & 23.1 & 23 \\
& \textbf{Llama3} & \rnl  & \textbf{71.8} & \textbf{72.6} & \textbf{71.7} &  \underline{75.3} & 84.9 & \textbf{79.8} & 194 & \underline{65.8} & 51 & 57.5 & 76 \\ \cmidrule(lr){2-14}
& \textbf{GPT-3.5} & \rnl + \cdiff & 65.6 & 61.5 & 62.2 &  75.8 & 58.1 & 65.8 & 132 & 47.8 & 67.3 & 55.9 & 138 \\
& \textbf{CodeLlama} & \rnl + \cdiff & 54.2 & 62.2 & 52.3  & 63.8 & \underline{94.2} & 76.1 & 254 & 37.5 & 6.1 & 10.5 & 16 \\
& \textbf{Llama3} & \rnl + \cdiff & 62.6 & 65.2 & 59.8 &  66.7 & 90.7 & 76.8 & 234 & 55.6 & 20.4 & 29.9 & 36 \\
\midrule 
\multirow{7}{*}{\rotatebox{90}{\paux}} & \textbf{GPT-3.5} & \rnl & 66.8 & 59.2 & 59.5 &  49.7 & 60.6 & 54.6 & 107 & 46.2 & \underline{76.3} & \underline{57.6} & 160 \\ 
& \textbf{CodeLlama} & \rnl  & \underline{71.0} & 71.7 & 70.1 &  73.2 & 87.7 & 79.8 & 205 & \textbf{67.2} & 43.9 & 53.1 & 64 \\
& \textbf{Llama3} & \rnl  & \underline{71.0} & \underline{71.9} & \underline{70.6} &  74.0 & 86.0 & \underline{79.6} & 200 & 65.7 & 46.9 & 54.8 & 70 \\ \cmidrule(lr){2-14}
& \textbf{GPT-3.5} & \rnl + \cdiff & 60.4 & 55.9 & 56.7 &  70.5 & 52.9 & 60.5 & 129 & 42.6 & 61.2 & 50.2 & 141 \\
& \textbf{CodeLlama} & \rnl + \cdiff & 47.0 & 62.2 & 55.7 &  64.1 & 92.4 & 75.7 & 248 & 40.9 & 9.2 & 15.0 & 22 \\
& \textbf{Llama3} & \rnl + \cdiff & 63.3 & 65.6 & 62.2 &  68.0 & 86.6 & 76.2 & 219 & 54.9 & 28.6 & 37.6 & 51 \\
\bottomrule
\multicolumn{14}{l}{\footnotesize The highest and second-highest results are in bold and underlined. \# represents the number of instances predicted in each class.} 
\end{tabular}
\end{table*}

%% file: tables/information_relevance_scores.tex
\begin{table}[t!]
\setlength{\tabcolsep}{4pt}
\centering
\caption{\label{tab:information_relevance_scores}Information and Relevance Scores for Different Training Sets on CodeReviewer model.}
\begin{tabular}{l|ll|ll}
\toprule
\multirow{2}{*}{\textbf{Training set}} & \multicolumn{2}{c|}{\textbf{Manual}} & \multicolumn{2}{c}{\textbf{Overall}}  \\
\cmidrule(lr){2-5} 
&  Information & Relevance & Information & Relevance \\
\midrule
\textsc{Original} & 3.44 &  2.48 & 3.69  & 2.36 \\
$\textsc{Cleaned}_{\textsc{GPT-3.5}}$ & \textbf{4.27}$^{*}$ & \textbf{2.77}$^{*}$ & \textbf{4.38}$^{*}$  & \textbf{2.63}$^{*}$\\
$\textsc{Cleaned}_{\textsc{Llama3}}$  & 4.08$^{*}$ & 2.72$^{*}$ & 3.85$^{*}$ &  2.38\\
\bottomrule
\multicolumn{5}{p{8.5cm}}{\footnotesize Information scores range from 1-5; relevance scores from 1-3. The $^*$ indicates that the score from cleaned model is significantly higher from the original model (p-value $<$0.05 using Wilcoxon signed rank tests).}
\end{tabular}
\end{table}